\documentclass[twocolumn,prb,showpacs,preprintnumbers,amsmath,amssymb,superscriptaddress]{revtex4}
\usepackage{graphicx}
\usepackage{epsfig}
\usepackage{bm}
\usepackage[usenames,dvipsnames]{color} 

\begin{document}

\title{Magnetic ordering induced by interladder coupling in the spin-1/2 Heisenberg two-leg ladder antiferromagnet C$_9$H$_{18}$N$_2$CuBr$_4$}

\author{Tao Hong}
\email[Electronic address: ]{hongt@ornl.gov}
\affiliation{Quantum Condensed Matter Division, Oak Ridge National Laboratory, Oak Ridge, Tennessee 37831, USA}
\author{K. P. Schmidt}
\author{K. Coester}
\affiliation{Lehrstuhl f$\ddot{u}$r Theoretische Physik I, TU Dortmund, D-44221 Dortmund, Germany}
\author{F. F. Awwadi}
\affiliation{Department of Chemistry, The University of Jordan, Amman 11942, Jordan}
\author{M. M. Turnbull}
\affiliation{Carlson School of Chemistry, Clark University, Worcester, Massachusetts 01610, USA}
\author{Y. Qiu}
\author{J. A. Rodriguez-Rivera}
\affiliation{National Institute of Standards and Technology, Gaithersburg, Maryland 20899, USA} \affiliation{Department of Materials Science and Engineering, University of Maryland, College Park, Maryland 20742, USA}
\author{M. Zhu}
\author{X. Ke}
\affiliation{Department of Physics and Astronomy, Michigan State University, East Lansing, Michigan 48824, USA}
\author{C. P. Aoyama}
\author{Y. Takano}
\affiliation{Department of Physics, University of Florida, Gainesville, Florida 32611, USA}
\author{Huibo Cao}
\author{W. Tian}
\author{J. Ma}
\affiliation{Quantum Condensed Matter Division, Oak Ridge National Laboratory, Oak Ridge, Tennessee 37831, USA}
\author{R. Custelcean}
\affiliation{Chemical Sciences Division, Oak Ridge National Laboratory, Oak Ridge, Tennessee 37831, USA}
\author{H.~D.~Zhou}
\affiliation{Department of Physics and Astronomy, University of Tennessee, Knoxville, Tennessee 37996, USA}
\author{M. Matsuda}
\affiliation{Quantum Condensed Matter Division, Oak Ridge National Laboratory, Oak Ridge, Tennessee 37831, USA}

\date{\today}

\begin{abstract}
We present specific-heat and neutron-scattering results for the \emph{S}=1/2 quantum antiferromagnet (dimethylammonium)(3,5-dimethylpyridinium)CuBr$_4$. The material orders magnetically at \emph{T}$_N$=1.99(2)\,K, and magnetic excitations are accompanied by an energy gap of 0.30(2) meV due to spin anisotropy. The system is best described as coupled two-leg spin-1/2 ladders with the leg exchange $J_{\rm leg}$=0.60(2)~meV, rung exchange $J_{\rm rung}$=0.64(9)~meV, interladder exchange $J_{\rm int}$=0.19(2)~meV, and an interaction-anisotropy parameter $\lambda$=0.93(2), according to inelastic neutron-scattering measurements. In contrast to most spin ladders reported to date, the material is a rare example in which the interladder coupling is very near the critical value required to drive the system to a N\'eel-ordered phase without an assistance of a magnetic field.
\end{abstract}



\maketitle

\section{Introduction}
Quantum spin-1/2 two-leg ladders have attracted much attention from both experimental and theoretical standpoints because of their rich physical properties. At zero magnetic field, inelastic neutron-scattering experiments on these materials\cite{Hong06:74,Notbohm07:98,Savici09:80,Schmidiger11:84,Schmidiger13:88} have provided comprehensive knowledge on one-magnon and unconventional two-magnon excitations\cite{Barnes47:93,Bouillot11:83,Normand11:83} predicted by theory. In magnetic fields,\cite{Chitra55:97,Giamarchi59:99} they exhibit novel quantum-critical behavior \cite{Sachdev99} such as Bose-Einstein condensation,\cite{Garlea07:98,Thielemann09:79} magnetic Bose glass,\cite{Hong10:81} and Tomonaga-Luttinger liquid phases.\cite{Thiel09:102,Hong10:105,Ninios12:108,Schmidiger12:108,Schmidiger13:111} In addition, a quantum phase transition is expected to occur from a quantum-disordered state to a magnetically-ordered state, as the strength of interladder couplings is increased.\cite{Troyer97:55,Normand97:56,Capriotii02:65}

To date, there have been very few detailed experimental studies of coupled spin ladders because of a scarcity of suitable model systems. Na$_2$Co$_2$(C$_2$O$_4$)$_3$(H$_2$O)$_2$ (Ref.~\onlinecite{Honda05:95}) and CaCu$_2$O$_3$ (Ref.~\onlinecite{Wolf05}) were previously identified as such systems on the basis of magnetization measurements. However, further investigation with neutron scattering\cite{Matsuda07:75} has found that Na$_2$Co$_2$(C$_2$O$_4$)$_3$(H$_2$O)$_2$ is a system of almost isolated dimers, with a singlet ground state, not of a ladder. In the case of CaCu$_2$O$_3$, neutron-diffraction work\cite{Kiryukhin01:63} has revealed an incommensurate spiral magnetic structure, which originates from frustrated interladder couplings, making the system more complicated than originally thought. Interladder couplings have also been reported in the two-leg spin-ladder compounds IPA-CuCl$_3$ (Refs.~\onlinecite{Masuda06:96,Fischer11:96}) and BiCu$_2$PO$_6$ (Ref.~\onlinecite{Plumb13:88}), but their ground states remain spin liquids owing to frustrated terms in their Hamiltonian. Three-dimensional (3D) long-range ordering (LRO) has been observed at $\sim$120 K in another ladder compound, LaCuO$_{2.5}$ (Ref.~\onlinecite{Kadono96:54}), but the details of the magnetic structure and the spin dynamics of this material are still unknown.

In the present paper, we report specific-heat measurements, and neutron-diffraction and inelastic neutron-scattering (INS) experiments on what we shall show to be a coupled ladder case, (dimethylammonium)(3,5-dimethylpyridinium)CuBr$_4$, which we call DLCB. This study reveals that a spin gap coexists in this material with a 3D LRO. We also determine main exchange interactions and interaction anisotropy from the measured dispersions both along and perpendicular to the ladder directions. The ratio $\alpha$=$J_{\rm int}$/$J_{\rm leg}$ of the interladder exchange, $J_{\rm int}$, to the leg exchange, $J_{\rm leg}$, indicates that DLCB is very close to a quantum critical point at which the LRO vanishes.

\begin{figure}[htbp]
\includegraphics[width=7.0cm,bbllx=50,bblly=65,bburx=555,bbury=675,angle=-90,clip=]{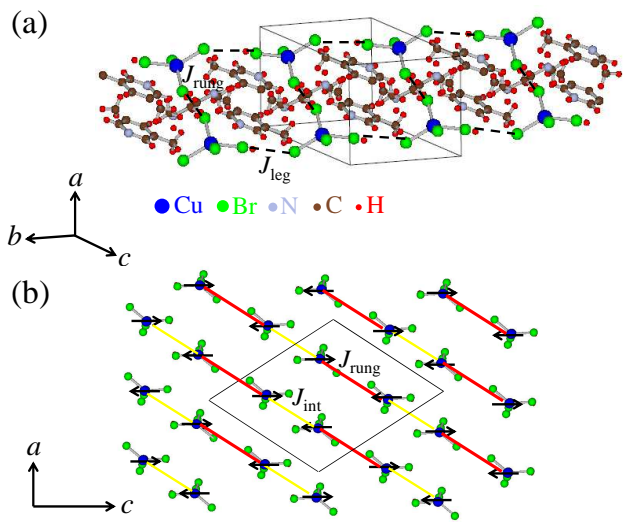}
\caption{(Color online) (a) Crystal structure of DLCB, in which the ladder chain extends along the $\bm{b}$ axis. Outlined is a nuclear unit cell. (b) Projection of CuBr$_4^{-2}$ tetrahedra onto a plane perpendicular to the $\bm{b}$ axis, showing proposed interladder couplings. Different lines stand for the different bonds. Red and yellow lines indicate the intraladder coupling $J_{\rm rung}$ and interladder coupling $J_{\rm int}$, respectively. Black arrows indicate the directions of the spins. Parallelogram is a projection of a magnetic unit cell.} \label{fig1}
\end{figure}

The crystal structure of DLCB is triclinic, space group $P\bar{1}$, and the lattice constants at 85 K are $a=7.459$ \AA, $b=8.270$ \AA, $c=13.720$ \AA, $\alpha=107.41^\circ$, $\beta=90.21^\circ$, and $\gamma=91.37^\circ$.\cite{Awwadi08:47} Nearest-neighbor and next-nearest-neighbor contacts between bromine atoms suggest that CuBr$_4^{-2}$ anions form two-leg ladders along the crystallographic $\bm{b}$ axis as shown in Fig.~\ref{fig1}(a).~$J_{\rm leg}$\,=\,0.685\,meV and a value of the rung exchange, $J_{\rm rung}$\,=\,0.351\,meV, have been obtained from magnetic susceptibility, which shows no evidence for LRO down to 2 K (Ref.~\onlinecite{Awwadi08:47}). But different from (Hpip)$_2$CuBr$_4$ (BPCB)\cite{Savici09:80,Thielemann09:79} and (2,3-dmpyH)$_2$CuBr$_4$ (DIMPY)\cite{Hong10:105,Schmidiger12:108,Shapiro07:952}, in which ladders are well separated from each other and thus interladder couplings ($\sim$$\mu$eV) are negligible, the interladder Cu-Cu distances in DLCB are comparable to or even shorter than intraladder Cu-Cu distances. Based on the crystal structure, we propose a minimal coupling between ladders as depicted in Fig.~\ref{fig1}(b), which shows the magnetic interactions between Cu$^{2+}$ ions in the crystallographic \emph{ac} plane. The red bond is $J_{\rm rung}$ and the yellow one is the nearest-neighbor $J_{\rm int}$ along the [10$\bar{1}$] direction in real space. As we shall show below, this two-dimensional (2D) model of coupled spin ladders fully accounts for experimentally observed magnon dispersions.

\section{Experimental Methods}
Single crystals of deuterated DLCB were grown according to the procedure described in Ref.~\onlinecite{Awwadi08:47}. The crystal structure was determined at 4 K on the four-cycle diffractometer (HB3A) at the High Flux Isotope Reactor (HFIR), Oak Ridge National Laboratory.

The magnetic neutron-diffraction measurements were made on a 0.2 g single crystal with a 0.4$^\circ$ mosaic spread, on a thermal neutron triple-axis spectrometer (HB1A) at the HFIR, with a neutron energy of 14.7 meV. A pyrolytic graphite (PG) (002) monochromator and analyzer were used together with horizontal collimation of 48$^\prime$-48$^\prime$-40$^\prime$-80$^\prime$. Contamination from higher-order beams was removed using PG filters. The sample was oriented horizontally in the (HK$\bar{\rm K}$) reciprocal-lattice plane and a continuous-flow helium-3 cryostat was used.

Inelastic neutron-scattering measurements were performed on a cold neutron triple-axis spectrometer (CTAX) at the HFIR and on a multi-angle crystal spectrometer (MACS)\cite{Rodriguez08:19} and a disk chopper time-of-flight spectrometer (DCS)\cite{Copley03} at the NIST Center for Neutron Research, on two co-aligned single crystals with a total mass of 1.0 g and a 1.0$^\circ$ mosaic spread. The sample was aligned in the either (HK$\bar{\rm H}$) or (HK0) scattering plane and standard helium-4 cryostats were used. At CTAX, the final neutron energy was set to either 3.5 or 5.0 meV by a PG (002) analyzer. The horizontal collimation was guide-open-80$^\prime$-open. At MACS, the final neutron energy was set to either 3.0 or 5.0 meV. Higher-order reflections were removed by a cooled beryllium filter placed between the sample and the analyzer. At DCS, a disk chopper was used to select a 167-Hz pulsed neutron beam with 3.27 meV. Reduction and analysis of the data were performed by using the software DAVE.\cite{Azuah09}

The specific heat was measured above 0.5 K by utilizing a commercial setup (Quantum Design, Physical Property Measurement System)\cite{ppms} and below 0.5 K in a home-made calorimeter\cite{Tsujii} in a dilution refrigerator.

\section{Experimental Results}
\begin{figure}[htbp]
\includegraphics[width=7.0cm,bbllx=0,bblly=15,bburx=550,bbury=720,angle=0,clip=]{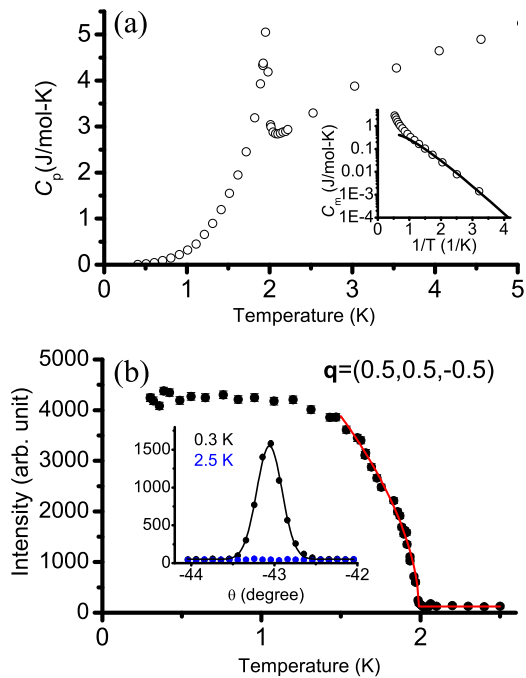}
\caption{(Color online) (a) Specific heat of DLCB. Inset: semilog plot of the magnetic component of specific heat, after subtraction of a phonon contribution, against $1/T$. The solid line is a fit showing exponential, activated behavior at low temperatures. (b) Background-subtracted neutron-peak intensity at $\bm{q}$=(0.5,0.5,$-$0.5) as a function of temperature. Error bars represent one standard deviation. The solid line is a fit to a power law as described in the text. Inset: rocking-curve scans through $\bm{q}$=(0.5,0.5,$-$0.5), made at \emph{T}=0.3 and 2.5 K. The solid line is a guide to the eye.} \label{fig2}
\end{figure}

To investigate the ground state of DLCB, we first performed specific-heat measurements, whose result is shown in Fig.~\ref{fig2}(a). The sharp anomaly at about 2.0 K indicates a phase transition to a LRO state. In addition, the exponential, activated behavior at low temperatures, shown in the inset to the figure, reveals the presence of a spin gap. By fitting the data to a formula for the specific heat of a one-dimensional $S$=1/2 gapped Heisenberg antiferromagnet in the low-temperature limit,\cite{Troyer94:50}

\begin{eqnarray}
C_{m} (T)\propto\left(\frac{\Delta}{k_B T}\right)^{3/2}\Delta e^{-\Delta/k_B T}, \label{eq:ct1}
\end{eqnarray}
we find the energy gap $\Delta$=0.29(2) meV. In a conventional $S$=1/2 two-leg spin-ladder system, the presence of a spin gap is tantamount to the ground state being quantum disordered. It might therefore seem surprising that a spin gap coexists with LRO in DLCB. We return later to this counterintuitive result.

To further examine the ordered ground state, we carried out single-crystal neutron diffraction measurements. At 0.3 K, we have collected in total 26 nuclear and 32 magnetic reflections, from which the magnetic propagation vector was found to be (0.5,0.5,0.5). The data were analyzed by the Rietveld method using the FULLPROF program.\cite{fullprof} The resulting collinear spin structure is depicted in Fig.~\ref{fig1}(b), with alternating moments pointing along the $\bm{c}^*$ axis in the reciprocal-lattice space. Not shown in the figure, the directions of the moments also alternate along the ladder. The size of the local moment is only 0.39(5)\,$\mu_B$ even at 0.3 K, much smaller than $\mu_B$, due to quantum fluctuations.

The inset to Fig.~\ref{fig2}(b) shows representative $\theta$ scans through $\bm{q}$=(0.5,0.5,$-$0.5) measured at 0.3 and 2.5 K. The scan at 0.3 K shows a peak, which disappears as the temperature is raised above about 2.0 K, thus confirming its magnetic origin. The temperature dependence of the intensity of the Bragg peak at $\bm{q}$=(0.5,0.5,$-$0.5) is plotted in Fig.~\ref{fig2}(b). A power-law fit of the form ($1-T/T_N$)$^{2\beta}$ gives a $T_N$ of 1.99(2)\,K, in good agreement with the specific-heat data, and the critical exponent $\beta$=0.28(2), which is smaller than $\beta$$\simeq$0.37 and 0.33 for the 3D Heisenberg and Ising universality classes, respectively.\cite{Peli}

\begin{figure}[htbp]
\includegraphics[width=8cm,bbllx=35,bblly=85,bburx=575,bbury=765,angle=0,clip=]{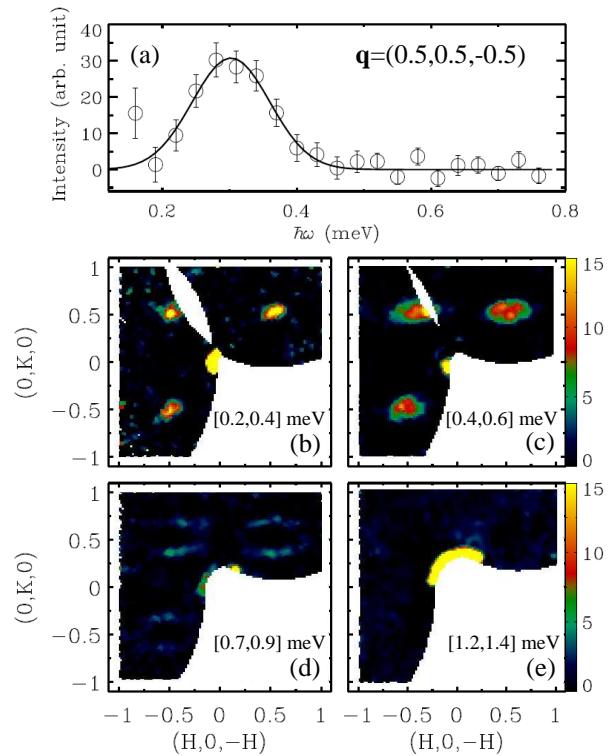}
\caption{(Color online) (a) Background-subtracted constant-\textbf{\emph{q}} scan in DLCB at the magnetic zone center (0.5,0.5,{$-$}0.5), measured at CTAX at \emph{T}=1.5 K. The solid line is a fit to a Gaussian profile convolved with the instrumental resolution function. Error bars represent one standard deviation. (b--e) Background-subtracted constant-energy slices measured at MACS at $T$=1.7 K for excitation energies of 0.2--0.4 meV, 0.4--0.6 meV, 0.7--0.9 meV, and 1.2--1.4 meV.} \label{fig3}
\end{figure}

INS was used to study the spin dynamics in DLCB. For all the results presented here, data taken at 10\,K with the same instrument configuration has been subtracted as a background. Figure~\ref{fig3}(a) shows a constant-$\bm{q}$ scan at the magnetic zone center (0.5,0.5,$-$0.5) at \emph{T}=1.5\,K. The instrumental-resolution-limited peak indicates a spin gap of 0.30(2) meV, in excellent agreement with the specific-heat result.

Figures~\ref{fig3}(b)--\ref{fig3}(e) show the evolution of the spin wave at 1.7 K in a few Brillouin zones (BZ) with increasing energy $\hbar\omega$. The magnetic intensity develops above the spin gap. As indicated by the ellipse-like shape of the intensity profile in Figs.~\ref{fig3}(b) and \ref{fig3}(c), the dispersion is weaker along [10$\bar{1}$], perpendicular to the ladder direction, than along the ladder direction [010]. The intensity disappears at $\hbar\omega$$\sim$0.8 meV for the dispersion perpendicular to the ladder and at about 1.3 meV for the dispersion along the ladder, as shown in Figs.~\ref{fig3}(d) and \ref{fig3}(e).

\begin{figure}[htbp]
\includegraphics[width=7.0cm,bbllx=20,bblly=30,bburx=500,bbury=790,angle=0,clip=]{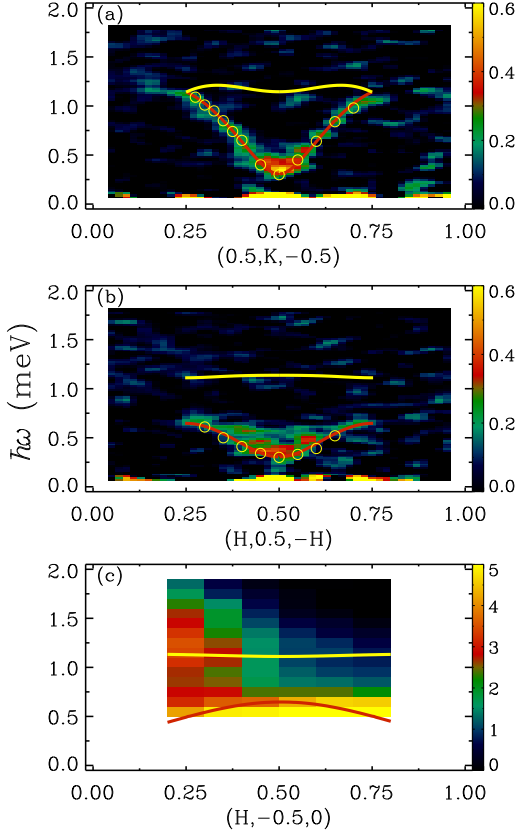}
\caption{(Color online) False-color map of the spin-wave spectra at $T$=1.5 K (a) along the ladders [010] and (b) perpendicular to the ladders along [10$\bar{1}$] measured at DCS, and (c) perpendicular to the ladders along [100] measured at MACS. Red and yellow lines are calculations for the acoustic and optical branches of one-magnon dispersion, respectively. Data shown as circles were obtained at CTAX.} \label{fig4}
\end{figure}

The spin-wave dispersions along these high-symmetry directions, [010] and [10$\bar{1}$], are presented in Figs.~\ref{fig4}(a) and \ref{fig4}(b), respectively. The bandwidths of the acoustic branch for these directions are 0.82(3) and 0.35(3) meV. As a result of the LRO, the magnetic unit cell is twice as large as the nuclear unit cell along the ladder direction. Consequently, the BZ is reduced to one half, and 0.25 and 0.75 become the zone boundaries. The observed spectrum termination due to this BZ folding is similar to that in the field-induced ordered phase of IPA-CuCl$_3$ (Ref.~\onlinecite{Zheludev07:76}). Furthermore, the presence of a flat optical branch as shown in Fig.~\ref{fig4}(c) arises from the alternation of $J$ along the [10$\bar{1}$] direction.

\section{Analysis and Discussion}
One peculiarity of the LRO in DLCB is that it does not produce a linear, gapless Nambu-Goldstone mode. This absence must arise from an inherent easy-axis anisotropy, which reduces the symmetry of the system to an axial one. In the ordered state, the spins form a collinear structure as shown in Fig.~\ref{fig1}(b), a structure which does not break the axial symmetry, hence the absence of a Nambu-Goldstone mode. This reasoning is borne out by detailed analysis of the spin-wave dispersions, as described below.

The simplest Hamiltonian that accounts for this scenario is
\begin{eqnarray} \label{ladeqn}
H &=&  \sum_{\gamma, \langle i,j\rangle}J_{\gamma} \left[ S^{\rm z}_{i}S^{\rm z}_{j}+\lambda\left( S^{\rm x}_{i} S^{\rm x}_{j}+S^{\rm y}_{i}S^{\rm y}_{j}\right)\right],
\end{eqnarray}
where the subscript $\gamma$ reads either `rung', `leg', or `int'---for $J_{\gamma}$ being the rung, leg, or interladder exchange constant---and $i$ and $j$ are nearest-neighbor lattice sites. The parameter $\lambda$ specifies an interaction anisotropy, with $\lambda$=0 and 1 being the limiting cases of Ising and Heisenberg interactions, respectively. We assume that $\lambda$ is the same for all three $J_{\gamma}$'s in order to minimize the number of fitting parameters to be determined from our spin-wave dispersion data.

To calculate the dispersion for this 2D model, we first perform a sublattice rotation, which transforms Eq.~(\ref{ladeqn}) to
\begin{equation} \label{ham_rotated}
 H =  \sum_{\gamma, \langle i,j\rangle} J_{\gamma} \left[-S^{\rm z}_{i} S^{\rm
 z}_{j}-\frac{\lambda}{2}\left(S^{\rm +}_{i}S^{\rm +}_{j}+S^{\rm -}_{i}S^{\rm -}_{j}\right)\right].
\end{equation}
We then introduce hardcore-boson operators $\hat{a}^{\dagger}_\nu$ and $\hat{a}^{\phantom{\dagger}}_\nu$ ($\hat{b}^{\dagger}_\nu$ and $\hat{b}^{\phantom{\dagger}}_\nu$), which create and annihilate a magnon at site $A$ ($B$) belonging to rung $\nu$ in the ferromagnetic reference state, obtaining the hardcore-boson Hamiltonian
\begin{eqnarray}\label{ham_bosons}
 \frac{\hat{H}}{\tilde{J}} &=& -\frac{N}{2}+\sum_{\nu}\left( \hat{n}_{\nu}^{(a)}+\hat{n}_{\nu}^{(b)}\right)\nonumber\\
                           && +\sum_\nu \left[ \hat{T}_{\nu,0}+\lambda\left( \hat{T}_{\nu,-2}+\hat{T}_{\nu,+2}\right)\right]\quad ,
\end{eqnarray}
where $\tilde{J}=J_{\rm leg}+(J_{\rm rung}+J_{\rm int})/2$, $N$ is the number of unit cells, $\hat{n}_{\nu}^{(a)}=\hat{a}^\dagger_\nu\hat{a}^{\phantom{\dagger}}_\nu$, and $\hat{n}_{\nu}^{(b)}=\hat{b}^\dagger_\nu\hat{b}^{\phantom{\dagger}}_\nu$. The sums are taken over all rungs. The operators $\hat{T}_{\nu,n}$, with $\hat{T}_{\nu,-2}=\hat{T}_{\nu,+2}^\dagger$, are given by
\begin{eqnarray}\label{T0}
         \hat{T}_{\nu,0}    &=& -x_{\rm rung} \hat{n}_{\nu}^{(a)}\hat{n}_{\nu}^{(b)}-x_{\rm int} \hat{n}_{\nu}^{(b)}\hat{n}_{\nu+\delta x}^{(a)}\nonumber\\
                          && -x_{\rm leg} \left( \hat{n}_{\nu}^{(a)}\hat{n}_{\nu+\delta y}^{(a)} + \hat{n}_{\nu}^{(b)}\hat{n}_{\nu+\delta y}^{(b)}\right)
\end{eqnarray}
and
 \begin{eqnarray}\label{T+2}
         \hat{T}_{\nu,+2}    &=& -x_{\rm rung} \hat{a}^\dagger_{\nu}\hat{b}^\dagger_{\nu}-x_{\rm int} \hat{b}^\dagger_{\nu}\hat{a}^\dagger_{\nu+\delta x}\nonumber\\
                           && -x_{\rm leg} \left( \hat{a}^\dagger_{\nu}\hat{a}^\dagger_{\nu+\delta y} + \hat{b}^\dagger_{\nu}\hat{b}^\dagger_{\nu+\delta y}\right) \quad ,
\end{eqnarray}
where $x_{\gamma}=J_{\gamma}/\tilde{J}$, and $\delta x$ ($\delta y$) is the distance between two neighboring rungs belonging to different ladders (the same ladder).

Finally, perturbative continuous unitary transformations\cite{Knetter2000,Knetter2003} map Eq.~(\ref{ham_bosons}) to an effective model, $H_{\rm eff}$, which conserves the number of magnons. The one-magnon sector $H^{\rm (1)}_{\rm eff}$, which is of our interest, can be simplified by Fourier transform to

\begin{equation}
 H^{\rm (1)}_{\rm eff} = \sum_{\bm{k}} \left( \omega_{\alpha}(\bm{k}) \alpha^\dagger_{{\bm{k}}} \alpha^{\phantom{\dagger}}_{\bm{k}}+  \omega_{\beta}(\bm{k}) \beta^\dagger_{{\bm{k}}} \beta^{\phantom{\dagger}}_{\bm{k}} \right) \quad ,
\end{equation}
where $\omega_{\alpha}(\bm{k})$ and $\omega_{\beta}(\bm{k})$ denote the two one-magnon branches stemming from the two-site unit cell. They are calculated as follows.

We first replace $x_\gamma$ in Eqs.~(\ref{T0}) and (\ref{T+2}) with $\tau x_\gamma$, so that at $\tau$=0, Eq.~(\ref{ham_bosons}) will contain only the first two terms. This is our unperturbed Hamiltonian. The perturbation expansion then gives $\omega_{\alpha}(\bm{k})$ and $\omega_{\beta}(\bm{k})$ as power series of $\tau$. We next express $\omega_{\alpha}^{2}(\bm{k})$ and $\omega_{\beta}^{2}(\bm{k})$ as Pad\'{e} approximants $P_{l}(\tau)/Q_{m}(\tau)$, where $P_{l}(\tau)$ and $Q_{m}(\tau)$ are power polynomials of order $l$ and $m$, respectively. The approximants are uniquely determined by choosing $l+m=n$, where $n$ is the order of the raw perturbation series.

The reason for casting $\omega_{\alpha}^{2}(\bm{k})$ and $\omega_{\beta}^{2}(\bm{k}),$ instead of $\omega_{\alpha}(\bm{k})$ and $\omega_{\beta}(\bm{k})$, as Pad\'{e} approximants is that the one-magnon energy gap \mbox{$\Delta=\omega(0,0)$} of a square-lattice antiferromagnet vanishes at $\lambda=1$ with a square-root singularity.\cite{Zheng91} In other words~$\Delta^2$ is a linear function~of $\lambda$ close to the critical point. This behavior of $\Delta^2$ is ensured by expressing $\omega_{\alpha}^{2}(\bm{k})$ as a Pad\'{e} approximant.

Finally, we set $\tau$=1 to obtain $\omega_{\alpha}^2(\bm{k})$ and $\omega_{\beta}^2(\bm{k})$ for the full Hamiltonian, Eq.~(\ref{ham_bosons}). We have found that perturbation expansion up to order $n=13$ results in
theoretical uncertainties that are well below the experimental error bars for all the data points shown in Fig.~\ref{fig4}.

To compare the calculation with the experimental data shown in Fig.~\ref{fig4}, we (i) shift the two spin-wave dispersions $\omega_{\alpha}(\bm{k})$ and $\omega_{\beta}(\bm{k})$ by $\bm{k}=(0,\pi)$ and (ii) fold the resulting dispersions into the reduced BZ in the $y$ direction, because the unit cell of the N\'eel-ordered state comprises four sites before the sublattice rotation. We choose the exchange constants $J_\gamma$ and $\lambda$ such that the calculated curves lie within experimental error bars. The results are in excellent agreement with the observed dispersions, as shown in Fig.~\ref{fig4}, yielding $J_{\rm leg}=0.60(2)$\,meV, $J_{\rm rung}=0.64(9)$\,meV, $J_{\rm int}=0.19(2)$\,meV, and $\lambda=0.93(2)$. The value of $J_{\rm leg}$ is somewhat smaller than 0.685 meV determined from magnetic susceptibility, whereas $J_{\rm rung}$ is much larger than 0.351 meV from the susceptibility.\cite{Awwadi08:47} The discrepancy is not surprising, given that the susceptibility analysis has assumed that $J_{\rm int}$ is negligible.

The exchange ratios $x$=$J_{\rm leg}/J_{\rm rung}$ and $\alpha$=$J_{\rm int}/J_{\rm leg}$ are~0.94(14) and~0.32(3), respectively. For Heisenberg $S$=1/2 coupled two-leg square ($x$=1) ladders, the quantum critical point between the spin-liquid phase and the N\'eel-ordered phase has been predicted to be at $\alpha_c$$\sim$0.3.\cite{Capriotii02:65} Since difference of the energy gaps of isolated spin ladders with $x$=0.94 and $x$=1 is as small as 0.01$J_{\rm rung}$, which is negligible,\cite{Hong10:105} we expect $\alpha_c$ for $x$=0.94 to be $\sim$0.3/0.94=0.32, which turns out to be the same as $\alpha$ for DLCB within our uncertainties. Thus DLCB is an ideal experimental realization of a coupled spin-ladder system in which the ground state becomes magnetically ordered with an $\alpha$ very close to the critical value. The weak Ising-like anisotropy ($\lambda$=0.93) prevents the gap from closing, while allowing---and to some extent even promoting---the ordering at finite temperature even if the interlayer coupling is absent.

\section{Conclusion}
In summary, we have carried out specific-heat and neutron-scattering measurements on DLCB to determine its spin Hamiltonian and the ground state. We have found a long-range magnetic order coexisting with a spin energy gap. An easy-axis anisotropy, which accounts for the coexistence, has also been identified. The measured magnetic dispersions are quantitatively consistent with a coupled $S$=1/2 two-leg spin-ladder model with ladder legs along the $\bm{b}$ direction.

\begin{acknowledgments}
TH thanks D. A. Tennant for a helpful discussion. We thank J.-H. Park and G. E. Jones for help with cryogenics. The work at the HFIR, Oak Ridge National Laboratory, was sponsored by the Division of Scientific User Facilities, Office of Basic Energy Science, US Department of Energy (DOE). The work at NIST utilized facilities supported by the NSF under Agreement Nos. DMR-9986442, -0086210, and -0454672. The National High Magnetic Field Laboratory (NHMFL), in which the specific-heat measurements were made, is supported by NSF Cooperative Agreement No. DMR-0654118, by the State of Florida, and by the DOE. XK acknowledges support from Michigan State University, CPA and YT acknowledge support by the NHMFL UCGP program, and HDZ acknowledges support from the JDRD program of the University of Tennessee.
\end{acknowledgments}

\end{document}